\documentclass[conference]{IEEEtran}
\IEEEoverridecommandlockouts
\usepackage{cite}
\usepackage{amsmath,amssymb,amsfonts}
\usepackage{algorithmic}
\usepackage{graphicx}
\usepackage[dvipsnames]{xcolor}
\usepackage[final]{microtype}
\usepackage[italic]{mathastext}
\usepackage{libertine}
\usepackage[T1]{fontenc}
\usepackage{textcomp}
\usepackage[varqu,varl]{zi4}
\usepackage[all]{nowidow}
\usepackage{fancyhdr}

\newcommand{\iiswcsubmissionnumber}{XX}
\fancypagestyle{firstpage}{
  \fancyhf{}
  
  \fancyhead[C]{\vspace{10pt}\normalsize{IISWC 2025 Submission
      \textbf{\#\iiswcsubmissionnumber} -- Confidential Draft -- Do
      NOT Distribute!!} \\\vspace{-25pt}} 
  \fancyfoot[C]{\thepage}
}

\begin{document}

\title{Characterizing Machine Learning Force Fields as Emerging Molecular Dynamics Workloads on Graphics Processing Units\\
%{\footnotesize \textsuperscript{*}}
%\thanks{The benchmark data with profiling examples available via : ....... %This work is funded by .........}
%}
\author{
\IEEEauthorblockN{
Udari De Alwis\textsuperscript{*}\IEEEauthorrefmark{3},
Benjamin E. Mayer\textsuperscript{*}\IEEEauthorrefmark{2}\thanks{This work was carried out while Benjamin Mayer was a Postdoctoral Researcher at imec.},
Tom J. Ashby\IEEEauthorrefmark{3},
Maria Barrera\IEEEauthorrefmark{3},
Timon Evenblij\IEEEauthorrefmark{3},
Joyjit Kundu\IEEEauthorrefmark{3}
}
\IEEEauthorblockA{\IEEEauthorrefmark{3}
IMEC, Leuven, Belgium\\
Email: \{udari.dealwis, tom.ashby, mariacecilia.barreramoreno, timon.evenblij, joyjit.kundu\}@imec.be
}
\IEEEauthorblockA{\IEEEauthorrefmark{2}
Email: benjamin.e.mayer@gmail.com
}

{\footnotesize \textsuperscript{*}}
\thanks{*Equal contributing authors}
%\thanks{The benchmark data with profiling examples available via : .......}
}

}

\maketitle
\begin{abstract}
Molecular dynamics (MD) simulates the time evolution of atomic systems governed by interatomic forces, and the fidelity of these simulations depends critically on the underlying force model. Classical force fields (CFFs) rely on fixed functional forms fitted to experimental or theoretical data, offering computational efficiency and broad applicability but limited accuracy in chemically diverse or reactive environments. In contrast, machine learning force fields (MLFFs) deliver near–quantum-chemical accuracy at molecular-mechanics cost by learning interatomic interactions directly from high-level electronic-structure data.

While MLFFs offer improved accuracy at a fraction of the cost of quantum methods, they introduce significant computational overhead, particularly in descriptor evaluation and neural network inference. These operations pose challenges for parallel hardware due to irregular memory access, minimum data reuse and inefficient kernel execution. 

This work investigates the hardware performance of such models using poly-alanine chains, a novel benchmark molecule system(s) with controllable input size, which used as performance evaluation test cases highlighting the computational bottlenecks of the graphical processor units when scaling out MLFF simulations. The analysis identifies key bottlenecks in descriptor and force computation, memory handling, highlighting the opportunities for improvements in the emerging area of MLFF based MD in drug discovery, that has received limited attention from a computer architecture perspective.
\end{abstract}


\section{Introduction}
\begin{figure*}[tb]
\centering
\centerline{\includegraphics[width=14cm]{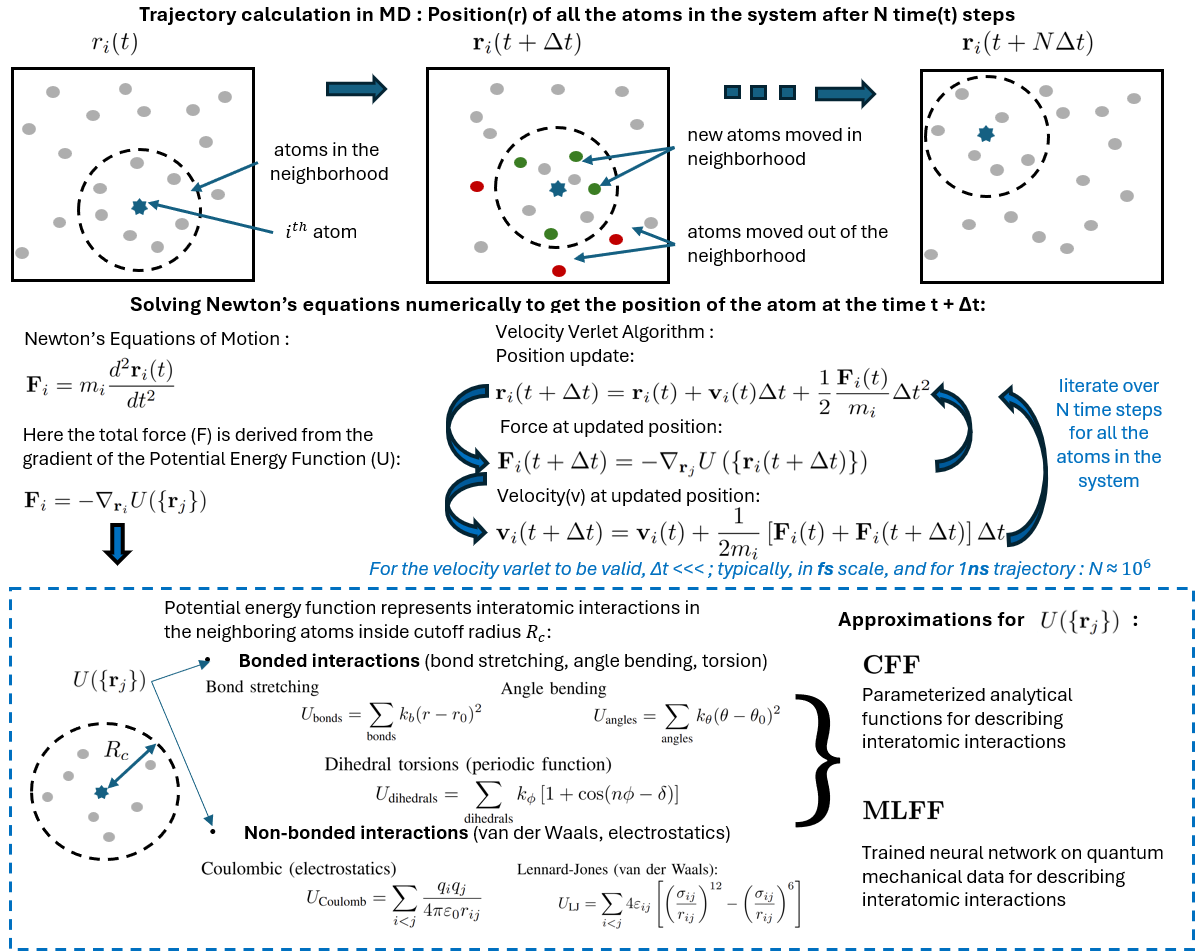}}
\caption{Trajectory calculation in Molecular Dynamics and Potential Energy function approximates for force calculation step in the trajectory (considering a fixed cutoff system excluding long range interactions).}
\label{md_trajectory}
\end{figure*}

% Molecular Dynamics (MD) simulations are a cornerstone workload in life sciences~\cite{Ahmed2023} and materials science~\cite{Chew2025}, capturing the temporal evolution of atomic systems via numerical integration of Newton’s equations of motion (Fig.~\ref{md_trajectory}). MD problem sizes range from $\sim$1k to $100$M atoms, with demonstrated parallel efficiency of $0.9$ on $65$k cores~\cite{Kutzner2025,Pll2020}. While quantum mechanics (QM) provides highest fidelity, it is intractable for large-scale systems. Classical force fields (CFFs), empirical functional forms parameterized against QM ~\cite{case2020amber,brooks2009charmm} enable scalable MD at reduced accuracy. Despite steady advances~\cite{Brini2020}, CFFs fail to generalize in critical applications such as drug binding affinity~\cite{SabansZariquiey2024,Chen2024} and complex material modeling~\cite{Rcken2024}.Machine Learning Force Fields (MLFFs), such as ANIxx \cite{ANI1_20217,ANI_2020,morado_2023}, represent a new generation of interatomic potentials that aim to overcome these limitatibons.
Molecular Dynamics (MD) is a widely used computational technique \cite{Ahmed2023,Chew2025,Hollingsworth2018} for simulating the temporal evolution of atomic and molecular systems by numerically solving Newton’s equations of motion (Fig.~\ref{md_trajectory}). The accuracy and computational efficiency of MD simulations are fundamentally determined by the Force Field (FF) used to model interatomic interactions. 

Classical Force Fields (CFFs), such as AMBER \cite{case2020amber} and CHARMM \cite{brooks2009charmm}, rely on empirical functional forms with parameters fitted to experimental and quantum mechanical data. Although CFFs offer substantial computational efficiency and broad applicability, their predictive accuracy is limited, particularly in systems exhibiting chemically reactive environments where fixed parameters may fail to generalize. Hence, accuracy critical application areas like computational drug discovery require a more accurate class of FF.

Machine Learning Force Fields (MLFFs), such as ANIxx \cite{ANI1_20217,ANI_2020,morado_2023}, represent a new generation of interatomic potentials that aim to overcome these limitations. These models learn potential energy surfaces directly from high-level Quantum Mechanical (QM) calculations, using descriptors to encode the local atomic neighborhood. Neural networks are then used to predict atomic contributions to total energy. This approach enables MLFFs to achieve high accuracy compared to CFFs. Yet the computational overhead incurred from MLFFs is significantly higher than that of CFFs. 

Although the scalability of MD with classical force fields (CFFs) is well studied\cite{Naderan_2021,Paverelli_2022,machado2022mdbench,Jones2022}, the computational overhead of Machine Learning FFs (MLFFs) remains poorly understood. In drug design, MLFFs often model parts of the system where accuracy is most crucial, due to the high computational overhead they incur, resulting in a hybrid workload that combines a well-characterized classical MD workload with a less understood ML workload. However, with advances in more efficient compute infrastructure, it is possible to expand the application of MLFFs to larger atomic systems, enabling more accurate predictions for drug discovery applications. Identifying compute bottlenecks in scaled-up MLFF simulations is crucial for designing future hardware. This work addresses this knowledge gap by analyzing MLFF performance when scaling molecular workloads on GPU architectures and offering optimization recommendations based on observed bottlenecks. To the authors best knowledge, this is one of the initial studies investigating the computer architectural perspective of MLFF workload scaling.

This paper contributes by: (a) designing variable-sized molecular structures chemically compatible with MLFFs, (b) analyzing multiple MLFF network architectures and atomic neighborhood embedding performance on GPUs, and (c) identifying workflow bottlenecks with profiling-based recommendations. Sections II–VI cover background, molecular system setup and profiling, performance analysis, bottleneck evaluation, and conclusions.
\section{Background}
\subsection{Molecular Dynamics in Drug Discovery and Machine Learning Force Fields (MLFFs)}
Standard MD integrates molecular mechanics via numerical schemes (Fig.~\ref{md_trajectory}). Forces split into short-range (bonded) and long-range (non-bonded) terms. A naive all-pairs calculation requires $\mathcal{O}(n^2)$ operations, but cutoff-based methods reduce this to $\mathcal{O}(n \log n)$ or $\mathcal{O}(n)$~\cite{Frenkel2001yk}. 
Biomolecular systems span $10^5$–$10^8$ atoms, with simulation performance measured in ns/day. Recent studies report 687, 116 and 35 ns/day for 82k, 53M, and 204M atoms, respectively~\cite{Kutzner2025}. Since integration uses fs timesteps~\cite{Hollingsworth2018}, reaching biologically relevant ms–s timescales requires months of GPU compute~\cite{cerebras}. Strong scaling is limited by kernel launch and communication overhead in long-range electrostatics. Specialized accelerators (Cerebras WSE~\cite{cerebras}, ANTON~\cite{anton3}, MDGRAPE~\cite{mdgrape}) mitigate these issues, but none address MLFF workloads.

MLFFs approximate potential energy surfaces (PES) and atomic forces with near-quantum accuracy, offering accuracy over classical force fields and \textit{ab initio} methods~\cite{behler2007generalized}. A key neural network architecture is the Behler--Parrinello neural network (BPNN), which expresses system energy as a sum of atomic contributions. Each atomic energy is predicted by a species-specific neural network using atom-centered symmetry functions (radial + angular), designed to enforce translational and rotational invariance~\cite{behler2007generalized, behler2011atom}. ANI2x extends the BPNN framework with modified symmetry functions and training on large datasets (ANI-1x, ANI-1ccx) containing millions of density functional theory (DFT) conformers at the $\omega$B97X/6-31G(d) level~\cite{smith2017ani}. This enables broad chemical generalization, including conformational sampling and protein-ligand modeling. Beyond ANI, MLFFs evolved from system-specific (e.g., sGDML) to general-purpose architectures such as TorchMD-Net~\cite{pelaez2024torchmdnet}, SchNet and DeepMD. %, and espaloma~\cite{schutt2017quantum}. 
These frameworks support biomolecular simulations, capturing complex interactions and thermodynamics for protein folding, binding, and drug discovery~\cite{unke2021machine}.

\section{Molecular Systems for MLFF Simulations and GPU Profiling}
\subsection{Setup of Simulation Workload}
\begin{figure}[tb]
\centerline{\includegraphics[width =\columnwidth]{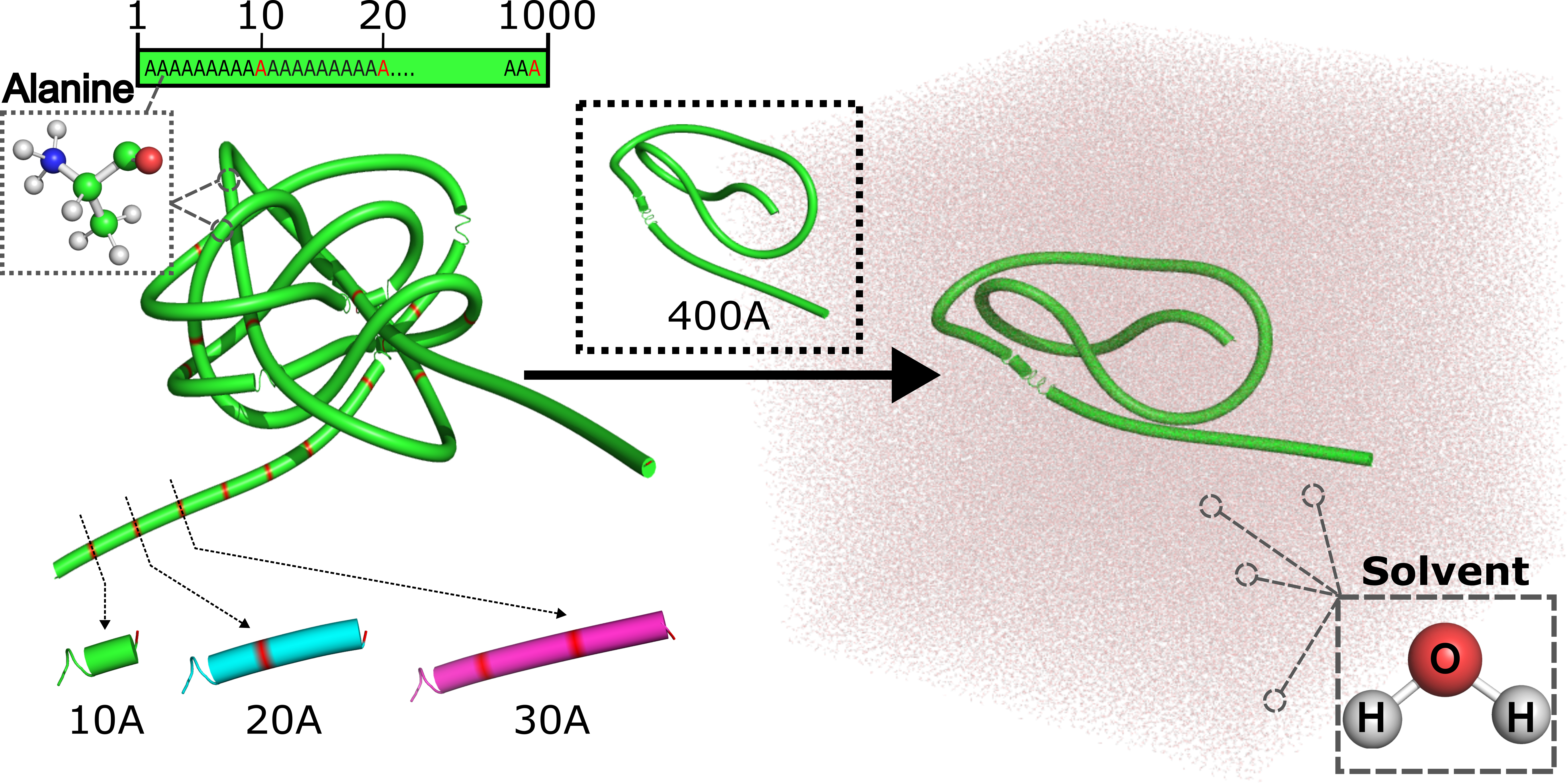}}
\caption{Visualization of benchmark system creation by incrementally extracting segments from a 1000-residue poly-alanine protein. Both the sequence and structure of the full-length protein are shown, with red highlights indicating cut points used to generate systems of varying sizes. Each resulting system is equilibrated in a water box.}
\label{system_prep}
\end{figure}

A molecular system that is chemically compatible and scalable, with controlled size, is an ideal candidate for evaluating the computational overheads incurred by MLFFs on GPUs. However, MLFFs, due to the nature of their training data, are typically not trained on large biomolecular systems. In contrast, CFFs are well established and scalable.
Since MLFFs are trained on DFT data for small molecules, their hypothetical utility lies in improving the dynamics of small molecules within larger systems. A plausible future computational workload involves applying MLFFs not only to the small molecule but also to the surrounding reaction area (the binding pocket) and the local environment, capturing their full interactions. Currently, MLFF training sets include only minimal peptide-like systems, such as alanine dipeptide, which, while slightly more complex than a single alanine residue, lack the structural complexity of real proteins that fold into 3D shapes. Although this use case extends beyond current training data, it remains relevant for performance studies. To stay close to the training domain while enabling scalability, poly-alanine, a homopolymer of alanine residues, was selected as the model system.
To have a sufficiently large system, a poly-alanine of 1000 residues, e.g., a sequence of 1000 alanines, was folded using AlphaFold2 \cite{Mirdita2022} \cite{Jumper2021} and used to create poly-alanine slices of (logarithmically) increasing size. This is visualized in Fig. \ref{system_prep}. As shown in Fig.\ref{system_prep}, a series of segments from the 1000-residue poly-alanine protein is generated to create a benchmark dataset. Segment sizes increase logarithmically, e.g., molecules with $[10, 20, \ldots, 100, 200, \ldots, 1000]$ amino acids.

To quantify the MLFF system's performance within the MD workflow, three configurations, CFFsys, MLFFsys, and CMLsys, were benchmarked using poly-alanine segments as input structures in OpenMM/Gromacs. CFFsys refers to the molecular system that uses the classical force field for both poly-alanine and the solvent (baseline). MLFFsys refers to the molecular system in which poly-alanine exists in a vacuum, with only the MLFF applied. CMLsys refers to the molecular system in which poly-alanine is simulated in water, with the MLFF for poly-alanine and the CFF for the rest of the system, including interactions between the solvent and the poly-alanine molecule. When adding the solute to the scaled-up poly-alanine residues, due to the folded protein's structure, the solute diameter remained almost constant from 200 to 900 residues. However, the number of atoms in the system increased by approximately a factor of 2.5 upon adding the last 100 residues, as the protein's shape changed significantly.

\subsection{GPU Profiler Setup and Data Acquisition}
The MD simulation is performed through many sequential time steps. (Fig.\ref{md_trajectory}). The time measurements were carried out over 1000 steps in the production run, with a 0.5 fs MD simulation resolution. For GPU resource utilization analysis, a single step in the production run (i.e., trajectory calculation) of the MD simulation is isolated (using the CUDA Profiler API in PyTorch with OpenMM and checkpoints in Gromacs) for profiling. The production step comprises both potential and force calculation steps. Both Nsight Systems (NSys)~\cite{nvidia_nsys} and Nsight Compute (NCU)~\cite{nvidia_ncu} were used to profile the application's overall hardware resource utilization and the individual kernels' resource utilization. The first few steps of the simulation are excluded to remove the setup overhead from the application execution in OpenMM/Gromacs and PyTorch compilation. Before profiling, energy-minimization steps are performed to ensure the molecular system's stability. 
In addition to running MLFFs in the main MD flow, the MLFFs (ANi2x,TorchMd-net) were profiled separately using the Pytorch profiler~\cite{pytorch_profiler}  along with NSys and NCU to quantify the isolated effect of the ML forward propagation and backpropagation paths using randomly generated atomic coordinates in varying system sizes. 

GPU experiments were performed on a compute node with an NVIDIA A100 80GB PCIe card for base experiments; other GPU experiments were performed on NVIDIA A100 40GB SXM4, NVIDIA H100 80GB HBM3, and Tesla V100 PCIe 32GB GPUs residing in the same cluster.
\section{MLFF Architecture Analysis}
ANI-2x and TorchMD-Net share a common architectural foundation rooted in locality, symmetry preservation, and atomic energy decomposition. Both represent molecular energies as sums of atom-centered contributions derived from local neighborhoods, encoded in a way that respects the translational, rotational, and permutational symmetries of physical systems. Deep neural architectures refine these local descriptors to capture complex many-body effects, while forces are obtained via automatic differentiation of the learned potential. These shared principles allow both models to provide quantum-accurate predictions at a computational cost suitable for large-scale molecular dynamics simulations.

\subsection{ANI2x}
ANI2x~\cite{ANI_2020} is an ensemble-based neural network potential designed to predict molecular energies and forces while preserving physical symmetries. Its framework consists of three main stages: (i) atomic environment embedding via Atomic Environment Vectors (AEVs), (ii) energy prediction through element-specific multilayer perceptrons, and (iii) force evaluation via automatic differentiation. 

\subsubsection{Atomic Environment Vector (AEV)}
Each atom is described by a fixed-length descriptor encoding its local environment within a cutoff radius $R_c \approx 5.1~\mathrm{\AA}$. The AEV consists of radial and angular components:
\begin{equation}
G^R_m(i) = \sum_{j \ne i} e^{-\eta (R_{ij} - R_s)^2} \cdot f_c(R_{ij}),
\label{radial_summetry_function}
\end{equation}
\begin{equation}
f_c(R_{ij}) = 
\begin{cases}
\frac{1}{2} \left[\cos\left(\frac{\pi R_{ij}}{R_c}\right) + 1\right], & R_{ij} \le R_c, \\
0, & R_{ij} > R_c,
\end{cases}
\label{cutoff_function}
\end{equation}
\begin{equation}
%\resizebox{\columnwidth}{!}{$
\begin{split}
G^{A}_m(i) = 2^{1-\zeta} \sum_{j \ne i} \sum_{k \ne i,j} 
\left[1 + \cos(\theta_{ijk} - \theta_s)\right]^\zeta 
\\ \cdot \exp\left[-\eta \left(\frac{R_{ij} + R_{ik}}{2} - R_s\right)^2\right] 
\cdot f_c(R_{ij}) \cdot f_c
%$}
\end{split}
\label{radial_summe_function}
\end{equation}
Radial symmetry functions $G_m^R$ encode pairwise distances with Gaussian basis functions modulated by a smooth cutoff, while angular functions $G_m^A$ encode triplet angles $\theta_{ijk}$ to capture local geometric arrangements. The AEV size grows with the number of element types and the number of radial and angular functions. For ANI2x, the resulting AEV has 1008 components.

\subsubsection{Neural Network Potentials (NNP)}
ANI2x employs element-specific neural network potentials to efficiently predict molecular energies and forces. The total molecular energy is expressed as a sum of atomic contributions:
\begin{equation}
E_T = \sum_{i=1}^{N} \mathcal{N}_{X_i} \left( \mathbf{G}_i^X \right),
\label{total_energy}
\end{equation}
where $N$ is the number of atoms, $X_i$ the atomic number of atom $i$, $\mathcal{N}_{X_i}$ an element-specific MLP, and $\mathbf{G}_i^X$ the atomic environment vector (AEV) representing the local chemical environment. Ensembles of MLPs per element improve prediction accuracy over single large networks. 
\subsubsection{Energy and Force Evaluation}
Forces are computed via automatic differentiation:
\begin{equation}
\mathbf{F}_i = -\frac{\partial E}{\partial \mathbf{r}_i}.
\end{equation}
During a forward pass, atomic coordinates are mapped through the ensemble MLPs to compute the total energy $E_T$. Forces are obtained by differentiating the energy with respect to atomic positions, propagating gradients through the network:
\begin{equation}
\frac{\partial E}{\partial \mathbf{r}_i} = 
\frac{\partial E}{\partial h_n} \cdot 
\frac{\partial h_n}{\partial h_{n-1}} \cdots 
\frac{\partial h_1}{\partial \mathbf{r}_i},
\end{equation}
or equivalently, using the AEV representation:
\begin{equation}
\mathbf{F}_j = - \frac{\partial E}{\partial \mathbf{r}_j} 
= - \sum_i \left( 
\frac{\partial E_i}{\partial \text{AEV}_i} \cdot 
\frac{\partial \text{AEV}_i}{\partial \mathbf{r}_j} 
\right).
\end{equation}
This ensures forces are consistent with the energy surface and preserves the physical symmetries of molecular systems.

\subsection{Ani2x Performance analysis }
\begin{figure}[tb]
\centerline{\includegraphics[width=\columnwidth]{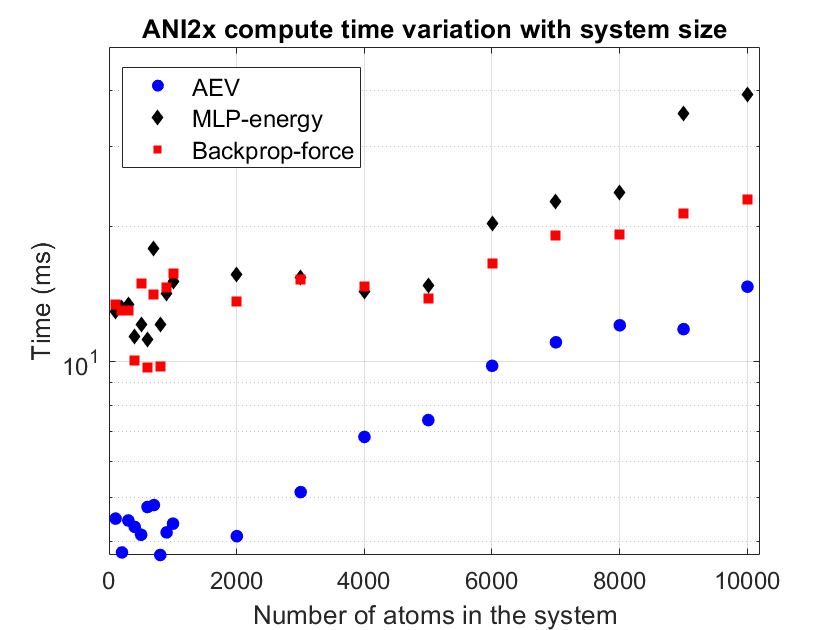}}
\caption{ANI2x potential/force compute time comparison with AEV compute time with varying system size.}
\label{AVE_neighbour_size}
\end{figure}

The performance of ANI2x as a function of system size was analyzed by measuring the computation time for three main stages: atomic environment vector (AEV) construction, neural network energy evaluation (MLP-energy), and force computation via backpropagation. The AEV calculation exhibits a roughly linear scaling with the number of atoms, reflecting its per-atom, neighbor-based nature, with minor fluctuations at small system sizes due to GPU underutilization (Fig.~\ref{AVE_neighbour_size}). In the AEV calculation step, the radial environment computation requires evaluating the function for each i,j atom pair in the cutoff region (without neighborhood lists). In contrast, the angular environment computation requires calculating the function for each i,j,and k triplet. For N atoms with an average number of M neighbors, the computational complexity adds up to $\mathcal{O}(N\times M)$ for the radial function. Analogous to this, the computation complexity adds up to $\mathcal{O}(N\times M \times M$) for the angular function.

In contrast, the MLP-energy evaluation shows a pronounced increase for large systems, particularly beyond 6,000 atoms, which can be attributed to dense linear operations and GPU memory overheads associated with large tensor manipulations. Force computation scales moderately with atom count because it requires propagating gradients through the AEV and MLP layers, with the number of neighbors per atom roughly constant. Overall, AEV computation remains computationally inexpensive, whereas MLP-energy evaluation dominates the simulation time for large molecules, and force calculations contribute a moderate but non-negligible fraction of the total cost. These observations indicate that ANI2x is efficient for small-to-medium systems but encounters memory and computational bottlenecks for very large molecular systems.

\begin{figure}[tb]
\centerline{\includegraphics[width=\columnwidth]{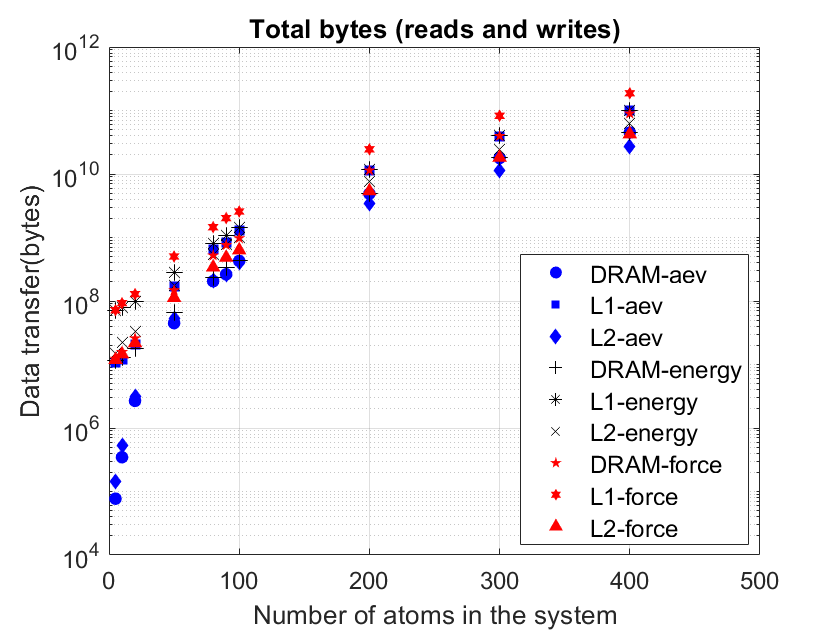}}
\caption{ANI2x AEV/potential/force memory transfers with varying neighborhood sizes.}
\label{memory_in_bytes}
\end{figure}

Furthermore, the memory requirement grows rapidly with the increase of  M(Fig.~\ref{memory_in_bytes}). In contrast to the much simpler GEMMs ordinarily used in MLPs, the computations here require storing intermediate results for the next step. Furthermore, a heavy burden is observed on the L2 cache from L1 misses arising from irregular memory access due to neighborhood changes, as well as from direct memory accesses without cache reuse. While the forward pass is relatively lightweight, the backward pass required for force computation demands significantly more memory and compute resources, especially for large systems with many atoms.
\begin{figure}[tb]
\centerline{\includegraphics[width=\columnwidth]{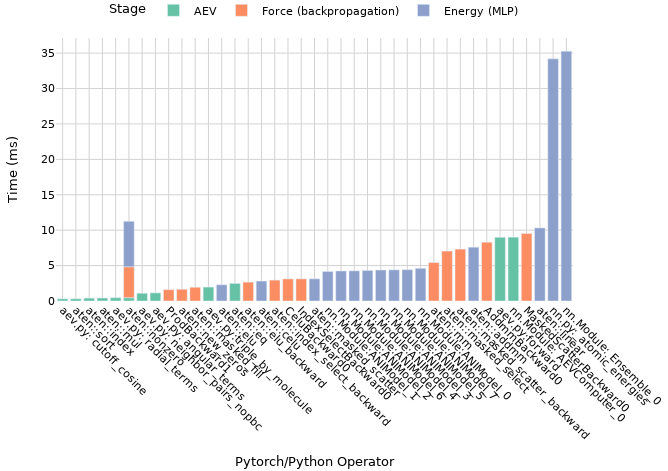}}
\caption{Combined Operator Timing: AEV vs Energy vs Force}
\label{ani_combined_kernels}
\end{figure}
% \subsection{Potential and Force Calculations in ANI2x MLFF }
\begin{figure}[tb]
 \centerline{\includegraphics[width=\columnwidth]{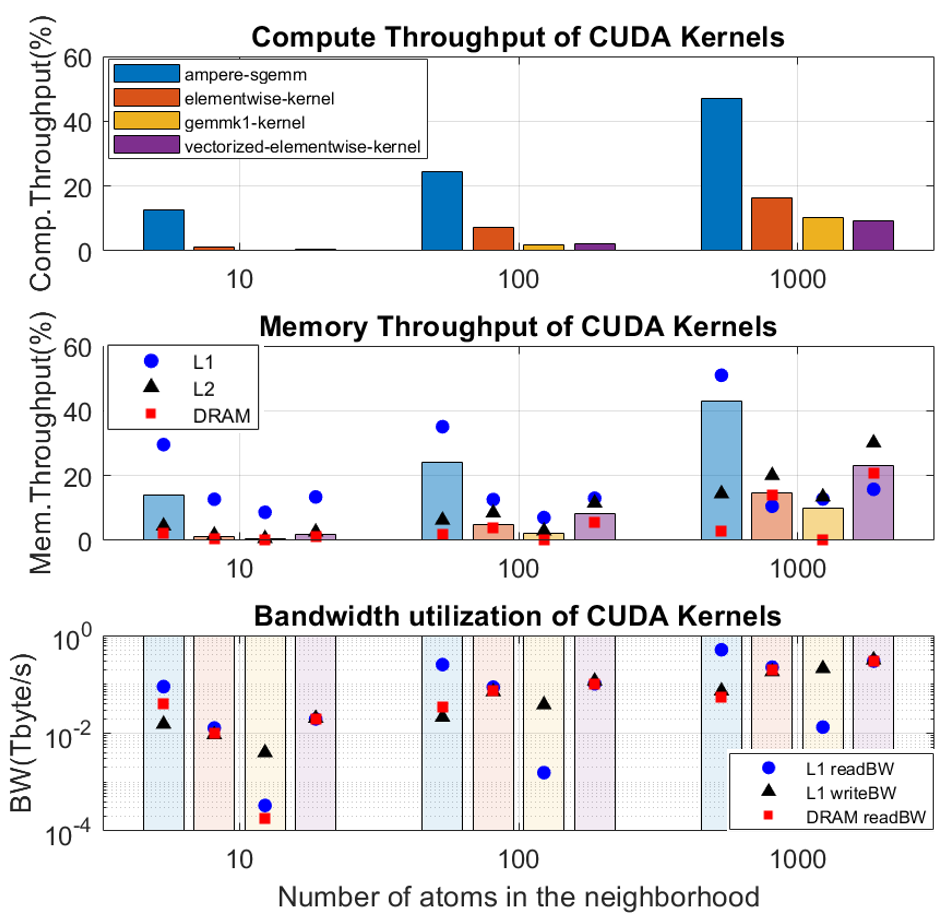}}
\caption{Relative Compute(SM) ,Memory Throughput and Bandwidth utilization of CUDA Kernels.}
\label{compute_memory_cuda}
\end{figure}
\subsubsection{Pytorch Function Analysis}
In the ANI2x neural network potential, computational hotspots differ across the AEV, energy, and force stages, reflecting the model's underlying algorithmic structure (Fig.~\ref{ani_combined_kernels}). During the AEV computation, the most time-consuming components are associated with neighbor-list generation and angular triplet enumeration, which rely heavily on memory-bound tensor indexing operations such as \texttt{aten::index}, \texttt{aten::index\_select}, and \texttt{aten::sort}. These kernels perform scattered memory accesses and masking operations, leading to low arithmetic intensity and reduced GPU utilization. In contrast, the energy evaluation stage is dominated by dense neural network inference, where fully connected layers implemented through \texttt{aten::linear} and \texttt{aten::addmm} account for the majority of the compute time. These operations are compute-bound matrix multiplications that scale directly with the number of atoms and the depth of the ANI sub-model ensemble. Finally, the force evaluation stage incurs the highest overall computational cost due to automatic differentiation. Backward-pass kernels such as \texttt{maskedScatterBackward}, \texttt{addmmBackward}, and \texttt{aten::maskedSelect} dominate runtime because they require both gradient propagation and reconstruction of intermediate values while performing irregular gather scatter operations. These gradient kernels exhibit poor memory locality and synchronization overheads, contributing to the substantial time spent in force calculation.

\subsubsection{CUDA Kernel Analysis}
Fig.~\ref{compute_memory_cuda} illustrates memory and compute throughput across a few kernel types which are widely used in the AEV, potential (forward propagation), and force (backward propagation) of the ANI2x MLFF (which accounts for up to 44\%-58\% total GPU runtime). Backpropagation through the AEV computation and multiple neural network layers triggers more CUDA kernel launches than the forward pass. According to NCU kernel profiling, the TorchANI implementation of the ANI2x energy calculation involves 77–81 CUDA kernel launches for neighborhood sizes ranging from 10 to 100 atoms. In comparison, the force computation for the same neighborhood sizes requires up to 96 CUDA kernel launches. These throughput matrices indicate a given kernel's effective utilization of the GPU hardware resources. A molecule with a few hundred atoms underutilizes GPU resources, as the kernel grids do not fully fill the devices' available resources, leading to fewer full waves across all streaming multiprocessors (SMs). Furthermore, the overall memory BW utilization falls below 1 Tbytes/s for the considered CUDA kernels, indicating underutilization of the hardware memory resources.
As the workload size increases, the kernel's compute and memory utilization increase, resulting in efficient hardware resource utilization and amortized kernel launch overheads. 
In the AEV calculation, each iteration involves computing atomic neighborhoods based on interatomic distances (eq.~\ref{cutoff_function}). When atoms are spatially arranged irregularly, this leads to non-uniform neighborhood sizes. Consequently, GPU threads access scattered memory addresses, resulting in uncoalesced memory accesses. This impairs cache efficiency and increases memory latency, reducing GPU performance.

\subsection{TorchMD-Net : Equivariant Transformer(ET) Architecture}
TorchMD-Net~\cite{torchmd_nET2022} is an equivariant neural network model designed for learning molecular potential energy surfaces and force fields with strong physical priors. Its architecture consists of three principal stages: (i) atomic embedding and environmental encoding, (ii) energy prediction, and (iii) force evaluation. The model ensures rotational, translational, and permutational symmetries that are fundamental to molecular physics.

\subsubsection{Embedding and Environmental Encoding}
Each atom, identified by its atomic number \( Z_i \), is initially mapped to a learned embedding:
\begin{equation}
    \mathbf{h}_i^{(0)} = \mathrm{Embed}(Z_i),
\end{equation}
providing an initial scalar feature representation. The next step incorporates local geometric information by constructing a neighbor list \( \mathcal{N}(i) \) containing all atoms \( j \) within a cutoff distance \( r_c \). The pairwise distance between atoms \( i \) and \( j \) is computed as:
\begin{equation}
    r_{ij} = \lVert \mathbf{r}_i - \mathbf{r}_j \rVert.
\end{equation}
This distance is expanded using radial basis functions (RBFs):
\begin{equation}
    \boldsymbol{\phi}_{ij} = \mathrm{RBF}(r_{ij}), \quad r_{ij} < r_c.
\end{equation}
TorchMD-Net represents each atom using two types of features: (i) scalar features \( \mathbf{h}_i \in \mathbb{R}^d \) (rotation-invariant) and (ii) vector features \( \mathbf{v}_i \in \mathbb{R}^{d \times 3} \) (rotation-equivariant). Under a rotation \( R \in SO(3) \), these features transform as:
\begin{equation}
    \mathbf{h}_i \rightarrow \mathbf{h}_i, \quad 
    \mathbf{v}_i \rightarrow R \mathbf{v}_i,
\end{equation}
ensuring the correct transformation behavior under rotation.

These features are updated through a sequence of ET layers, each of which performs attention-based message passing with geometric terms. The scalar features are updated as:
\begin{equation}
    \mathbf{h}_i' = \sum_{j \in \mathcal{N}(i)} \alpha_{ij} W_h\, \mathbf{h}_j,
\end{equation}
and the vector features as:
\begin{equation}
    \mathbf{v}_i' = \sum_{j \in \mathcal{N}(i)} \alpha_{ij} 
        W_v\!\left(\mathbf{v}_j + \mathbf{r}_{ij} \otimes \mathbf{h}_j\right),
\end{equation}
where \( \alpha_{ij} \) represents the distance modulated attention weights.
\subsubsection{Energy Prediction}
After \( L \) equivariant layers, each atom possesses refined scalar and vector features \( (\mathbf{h}_i^{(L)}, \mathbf{v}_i^{(L)}) \). Since potential energy is rotationally invariant, the vector features contribute through their magnitudes:
\begin{equation}
    \tilde{v}_i = \lVert \mathbf{v}_i^{(L)} \rVert.
\end{equation}
A lightweight neural network then maps these features to atomic energy contributions:
\begin{equation}
    E_i = f_\theta\big(\mathbf{h}_i^{(L)}, \tilde{v}_i\big),
\end{equation}
and the total energy is the sum of atomic contributions:
\begin{equation}
    E = \sum_{i=1}^{N} E_i.
\end{equation}

\subsubsection{Force Evaluation}
The forces acting on atoms are computed as the negative gradient of the total energy with respect to the atomic coordinates:
\begin{equation}
    \mathbf{F}_i = -\frac{\partial E}{\partial \mathbf{r}_i}.
\end{equation}

\subsection{Torchmd-NET Performance analysis}
\begin{figure}[tb]
 \centerline{\includegraphics[width=\columnwidth]{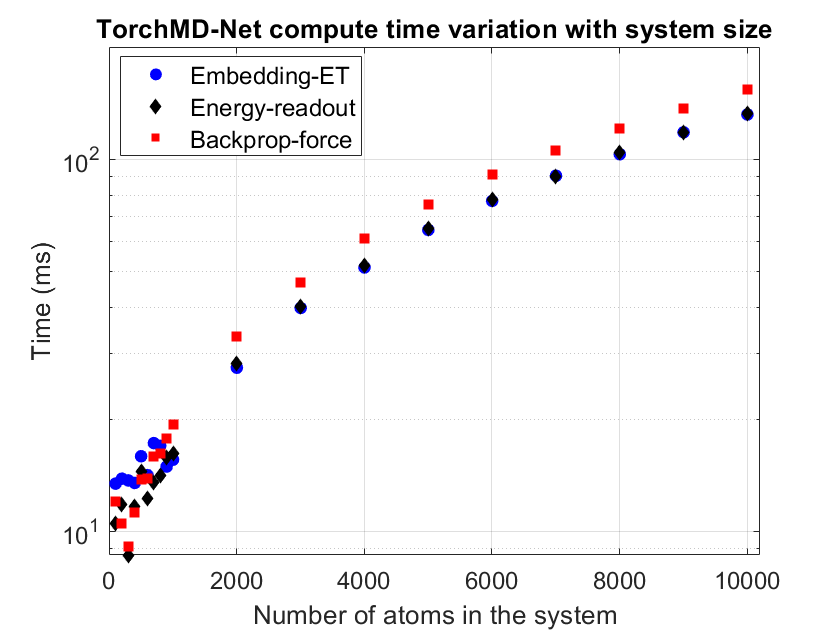}}
\caption{TorchMD-Net potential/force compute time comparison with Neighborhood Embedding compute
time with varying system size.}
\label{torchmdnet_layers}
\end{figure}

In TorchMD-Net, the dominant computational cost arises from the
equivariant multi-head attention mechanism employed within each layer of the ET. For a molecular system consisting of $N$ atoms with an average of $k$ neighbors per atom, the attention mechanism evaluates message-passing operations over approximately $Nk$ edges, with each edge processed across $H$ attention heads and $C$ hidden channels. These computations involve both scalar features and vector features, the latter requiring vector projections, directional normalization, and outer product based updates. As a result, the per layer computational complexity scales as ($\mathcal{O}(N k H C)$), making the vector equivariant attention kernels the primary bottleneck during GPU execution. Profiling indicates that $45$--$70\%$ of the total time is spent in these attention and message-passing operations (Fig.~\ref{torchmdnet_layers}). For small molecular systems (approximately 100–1000 atoms), the execution times of the ET, energy readout, and force computation stages remain relatively stable, exhibiting only minor fluctuations. This behavior indicates that fixed overheads, such as kernel launches, graph construction, and neighbor list initialization, dominate the computational cost in this regime. As the number of atoms increases beyond approximately 2000, the runtime of all three stages grows nearly linearly with system size, reflecting the increased cost of neighborhood-based message passing, equivariant attention operations, and scatter–reduce kernels. Among the three stages, force computation becomes the most computationally expensive for large systems due to the backward pass through both the ET representation and the energy readout layers.

\begin{figure}[tb]
 \centerline{\includegraphics[width=\columnwidth]{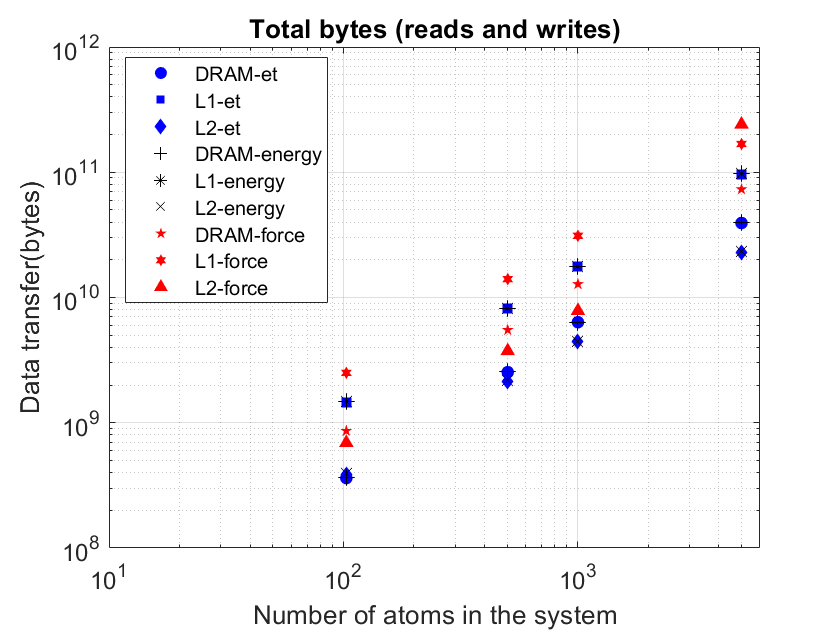}}
\caption{Torchmd-NET EMbedded-ET/potential/force memory transfers with varying neighbor-
hood sizes}
\label{torchmdnet_mem}
\end{figure}

The memory access statistics for TorchMD-Net exhibit a pronounced scaling trend with increasing system size (Fig.~\ref{torchmdnet_mem}). The DRAM, L2, and L1 traffic increase rapidly as the number of atoms grows, indicating a strong dependence on neighborhood size and message passing operations. The embedded ET and energy stages show nearly identical memory access patterns, reflecting their shared reliance on equivariant attention, scatter reduce operations, and feature aggregation. In contrast, the force computation stage generates substantially higher memory traffic across all cache levels, particularly in DRAM and L2, due to the backward pass through the full computational graph and repeated access to intermediate activations.
\begin{figure}[tb]
 \centerline{\includegraphics[width=\columnwidth]{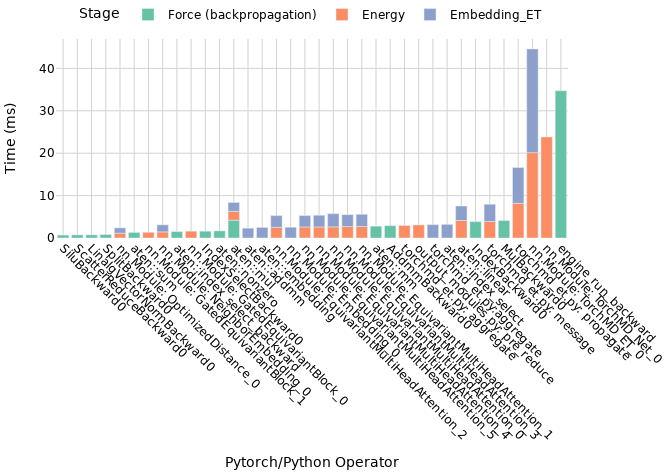}}
\caption{Combined Operator Timing: Embedded-ET vs Energy vs Force.}
\label{torchmdnet_ops}
\end{figure}
\subsubsection{Pytorch Function Analysis}
In TorchMD-Net, the dominant computational costs arise from the message-passing and equivariant attention mechanisms that underlie the model architecture(Fig.~\ref{torchmdnet_ops}). Within the embedded-ET stage, most of the runtime is spent inside the \texttt{TorchMD\_ET} module, where message passing, propagation, and aggregation constitute the principal bottlenecks. These operations rely heavily on irregular gather/scatter patterns, as evidenced by the substantial time spent in \texttt{aten::index\_select}, \texttt{scatter}, and \texttt{aggregate} routines. The multi-head equivariant attention blocks apply learned rotations and tensor projections to feature vectors, introducing additional linear layers and tensor reshaping costs. The message function itself incurs dense linear transformations (\texttt{aten::linear}) but remains secondary to memory-bound indexing and neighborhood aggregation overheads. In the energy-prediction stage, the same attention and message-passing kernels reappear, with dense linear algebra (\texttt{aten::linear}; \texttt{aten::addmm}) becoming more pronounced due to deeper transformations prior to readout. The force stage is dominated by the backward pass of these memory-intensive kernels, where autograd must replay all neighborhood-based indexing operations and linear transformations. As a result, gradient kernels such as \texttt{MulBackward0}, \texttt{AddmmBackward0}, and \texttt{IndexBackward0} account for most of the time.
\subsubsection{CUDA Kernel Analysis}
\begin{figure}[tb]
 \centerline{\includegraphics[width=\columnwidth]{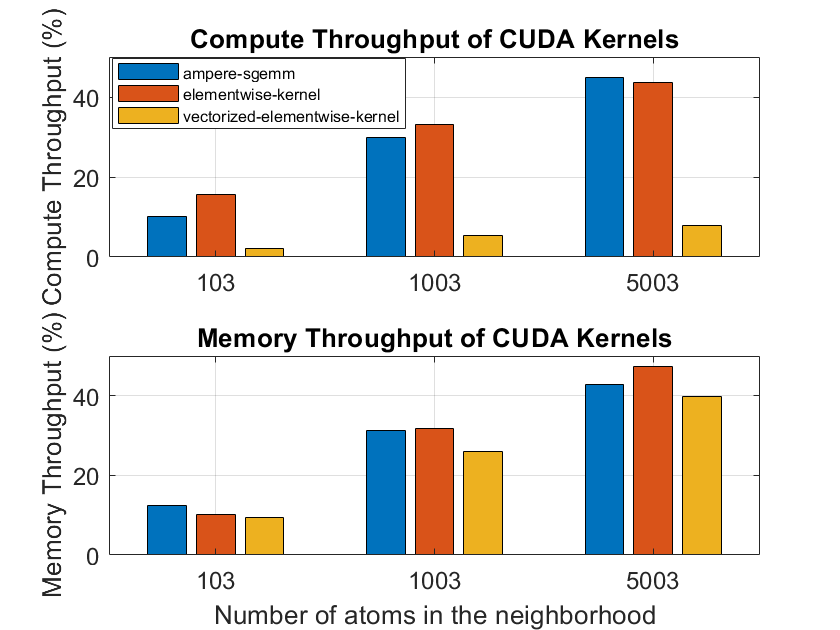}}
\caption{Relative Compute(SM),Memory Throughput of CUDA Kernels.}
\label{torchmdnet_sm_mem_util}
\end{figure}

For small molecular systems , CUDA kernels such as \texttt{ampere\_sgemm}, \texttt{elementwise\_kernel}, and \texttt{vectorized\_elementwise\_kernel} which accounts up to $57\%$ of runtime, operate on relatively small tensors, resulting in limited thread-level parallelism and low streaming multiprocessor utilization (Fig.~\ref{torchmdnet_sm_mem_util}). In this regime, kernel launch overheads and memory access latency dominate execution, leading to poor utilization of both compute and memory resources. As the atom count increases, the size of intermediate tensors associated with equivariant message passing, attention projections, and linear transformations grows substantially, enabling higher warp occupancy and more effective use of GPU execution units. This leads to a marked increase in compute throughput for \texttt{ampere\_sgemm} kernels. In parallel, memory throughput improves as larger workloads exhibit increased spatial locality and cache reuse, allowing \texttt{elementwise} and \texttt{vectorized\_elementwise} kernels to better amortize global memory access costs. Nevertheless, throughput remains significantly below peak hardware capabilities.
\section{MLFF Performance Analysis in the MD Trajectory Calculation }
\subsection{Comparison of CFF and MLFF Performances in GPUs}
\begin{figure}[tb]
\centerline{\includegraphics[width=\columnwidth]{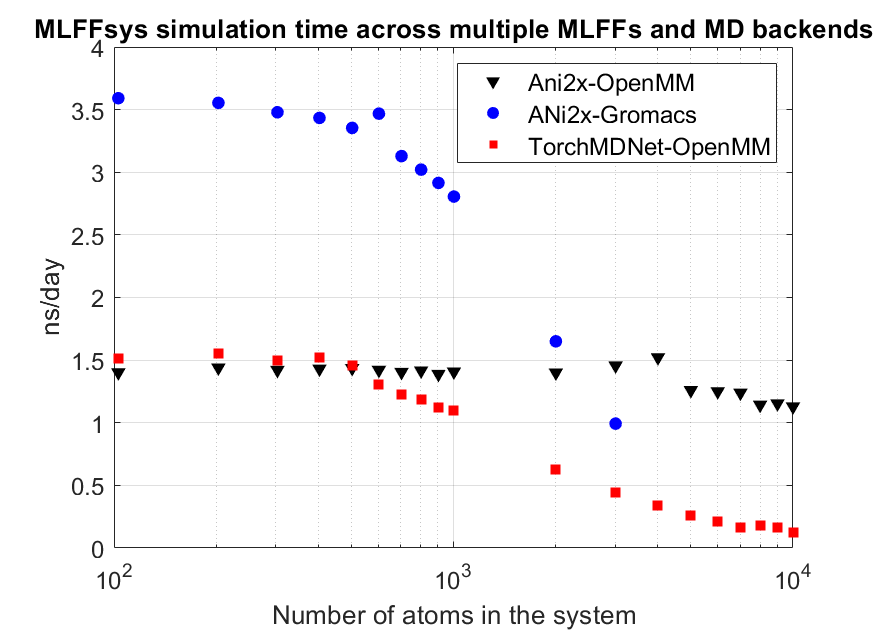}}
\caption{MLsys systems simulation times in MD production steps across varying workloads,MLFFs and MD back ends (simulated in single A100 80GB node).}
\label{ns_per_day_mlffsys}
\end{figure}

\begin{figure}[tb]
\centerline{\includegraphics[width=\columnwidth]{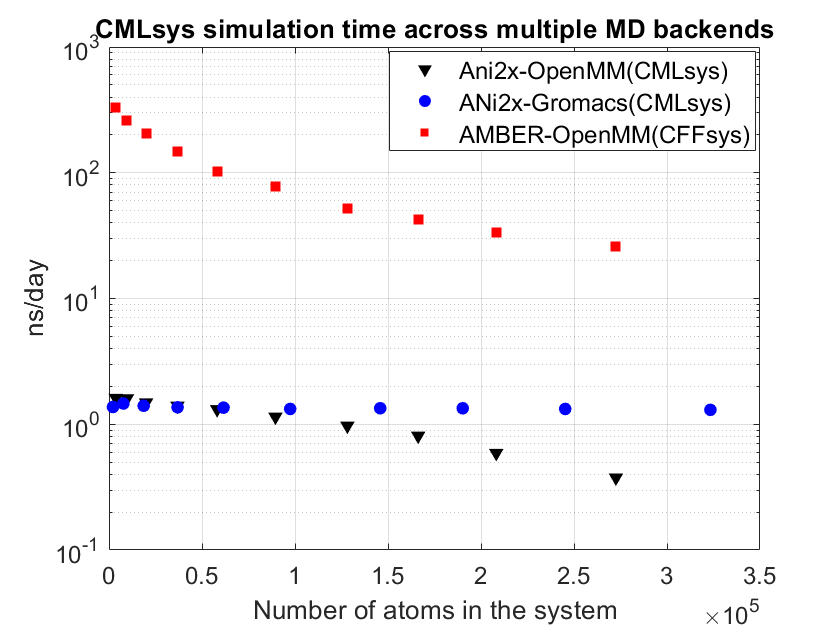}}
\caption{CMLsys systems simulation times in 1000 MD production steps across varying workloads (simulated in single A100 80GB node).}
\label{ns_per_day_cmlsys}
\end{figure}

NNPot~\cite{gromacs_nnpot} based Gromacs integrated ANi2x has slightly better performance in comparison to \texttt{openmm--ml}~\cite{OpenMM_ML} integrated Ani2x in OpenMM (Fig.~\ref{ns_per_day_mlffsys}). Yet the scaling of the workload demonstrate a steady decrease in the efficiency in Gromacs. Torchmd--Net is integrated in OpenMM using \texttt{TorchForce}~\cite{doerr2020torchmd} object. Even though Torchmd--Net has slightly better efficiency for the smaller atom numbers, the performance degrades steadily after 700 atoms.
Fig.~\ref{ns_per_day_cmlsys} summarizes the experimental results over a varying workload CMLsys, where ANI2x(MLFF) applied to the molecule while AMBER(CFF) applied to the solvent.Here ns/day is used as a measure of the compute efficiency. 
The CMLsys system requires 230x - 100x more compute time for a system size of 2000 - 35000 atoms. When the system scales to 145000 - 600000 atoms, the time difference is reduced to 75x-15x. When comparing ANi2x MLsys (Fig.~\ref{ns_per_day_mlffsys}) and CMLsys consume almost the same amount of time for the corresponding workload sizes, which indicates that the major contribution to the compute time comes from the MLFF computation. From the atomic interactions types, non-bonded interactions (Van der Waals and Coulombic interactions (Fig.~\ref{md_trajectory})) comprise a high computation overhead as this accounts for interactions between all the atoms in the system. When summing up the number of operations considering the analytical CFFs (equations listed in Fig.~\ref{md_trajectory}), the FLOPS accounts up to around 25 \textemdash 30 FLOPs per atom pair (i.e, for N atoms each having M neighbors, total short-range non-bonded $FLOPs \approx 25(N·M/2)$. For the Ani2x MLFF calculation, AEV layer accounts up to $\approx 16M + 32(M²/2)$ for a neighborhood having M pairs, while considering a single MLP for H, the multiply add operations adds up to $\approx 650k$ operations for single H atom. Comparing the same alanine dipeptide molecule in the two systems, CFF's operations add up to 8k while MLFF's sum up to 10M, which results in the observed 200x overhead even with highly parallel execution in MLPs.

\subsection{GPU Performance with Increased Workloads across multiple platforms}

When integrating ANI2x into OpenMM/Gromacs for molecular dynamics simulations, performance scaling across multiple GPUs is limited due to the lack of domain decomposition support within the OpenMM/Gromacs frameworks for the MLFFs unlike the CFF cases. This lack of domain decomposition prevents the ability to test the scalability of the MLFF performance across multiple GPU with their corresponding MD back ends. This is a future development necessary in the future release of the MD backend for the efficient use of MLFFs in larger system sizes.

\begin{figure}[tb]
\centerline{\includegraphics[width=\columnwidth]{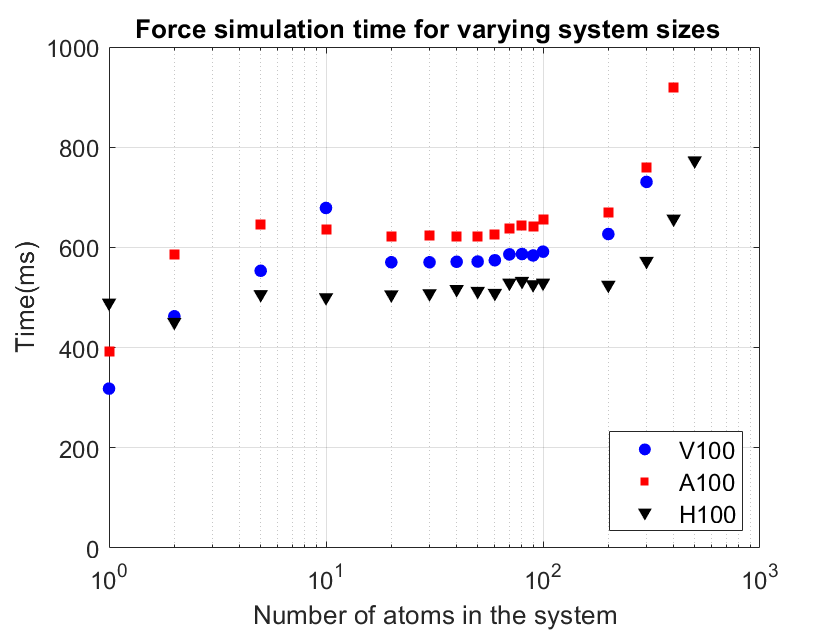}}
\caption{ANI2x force calculation with varying system sizes in TorchANI in varying GPU hardware platforms.}
\label{other_gpus}
\end{figure}

When benchmarking ANI2x force calculations across different GPU architectures (Fig.~\ref{other_gpus}), it was observed that the V100 outperforms the A100 for almost all system sizes. At the same time, the H100 consistently delivers superior performance across both small and large systems. The V100, with not so optimized scheduling model contributes better performance than A100 for MLFF workloads as MLFFs are dominated by large numbers of short-lived, low-arithmetic-intensity CUDA kernels rather than fused-long, throughput-bound kernels which, the more sophisticated scheduling in A100 catered for. The A100’s increased resources, such as more SMs and larger caches, are not fully utilized at low workloads, leading to suboptimal performance. In contrast, the H100 incorporates architectural advancements, including reduced kernel launch latency, and enhanced latency-hiding capabilities, enabling it to efficiently handle both small and large workloads.

\subsection{AEV Acceleration in GPUs}
\begin{figure}[tb]
\centerline{\includegraphics[width=\columnwidth]{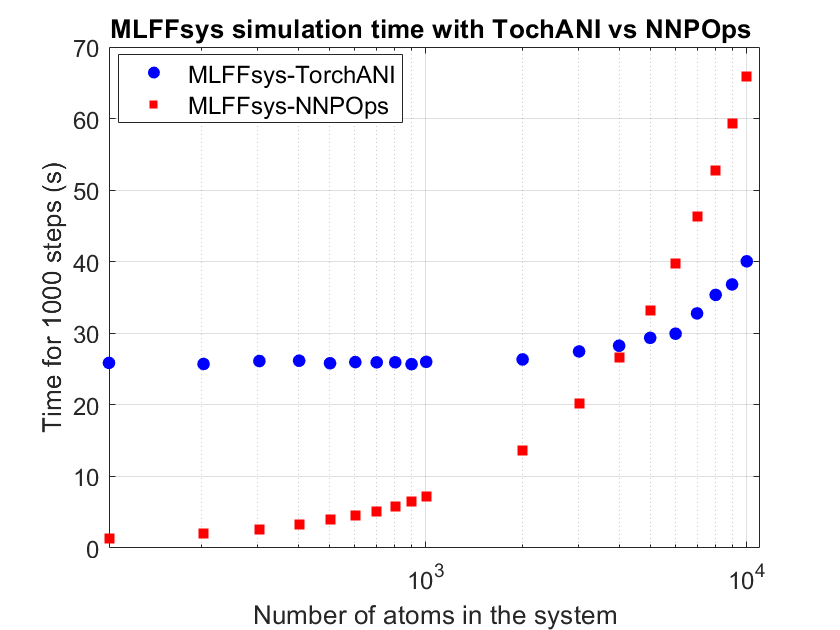}}
\caption{MLFFsys simulation time for 1000 steps for TorchANI vs NNPOps AEV accelerations).}
\label{mlsys_torchani_nnpops}
\end{figure}

NNPOps~\cite{NNPOps} accelerates AEV computation in ANI by replacing PyTorch-native tensor ops with fused CUDA kernels. TorchANI performs AEVs using standard PyTorch operations, resulting in $60-65$ CUDA kernel launches per AEV. While flexible, this leads to overhead from many small kernels, inefficient memory use, and poor GPU utilization, especially at neighborhood sizes of $50-100$ atoms typical in MD. TorchANI’s kernel-launch overhead can exceed actual compute for small neighborhoods, improving only gradually as neighbor size increases. NNPOps consolidates steps into four fused kernels, keeping intermediates in registers/shared memory and reducing launch overhead. NSys stats show $20\%-60\%$ CUDA API time in \texttt{cudaMallocManaged}, indicating unified memory overhead. NNPOps further applies spatial binning and optimized neighbor lists, yielding nearly $100×$ faster AEV evaluation, and $17-20×$ speedup including energy/force computation. NCU profiling shows radial/angular kernels are compute-bound on A100 for neighborhood sizes $10-100$, though small neighborhoods underutilize SMs. Coupled with OpenMM (Fig.~\ref{mlsys_torchani_nnpops}), NNPOps excels for small systems due to latency-optimized kernels, but scales poorly at larger sizes.
\subsection{Possible Hardware Software Optimizations}
Both ANI2x and TorchMD-Net with the OpenMM/Gromacs backend currently lacks multi-GPU support because it lacks domain decomposition for ML potentials. Spatial or atom-wise partitioning could enable data-parallel inference across GPUs enabling performance improvements. In ANI2x, each AEV consist of 1008 floats, plus intermediary activations and coordinate gradients required for backward paths. For large systems, this pushes GPU memory and hurts batch size. Mixed precision training/inference, AEV compression (e.g., using fewer symmetry functions or attention-based descriptors), or sparse representations could be a probable method in improving the memory burden of these MLFFs. Custom CUDA implementations (similar to NNPOps) and fused kernels can reduce launch latency and overlapping memory/compute stalls.
TorchMD-Net, particularly in its Equivariant Transformer embedding and message-passing stages, exhibits numerous small CUDA kernel launches and uncoalesced memory accesses, which result in suboptimal GPU utilization and high latency. Several strategies can be tried to mitigate these bottlenecks. Kernel and operator fusion can combine multiple small operations into a single CUDA kernel, thereby reducing launch overhead and increasing arithmetic intensity. Memory access patterns can be optimized by reordering atom indices, employing contiguous neighbor lists, or using compressed sparse representations, improving memory coalescing and throughput. Furthermore, attention and message-passing layers could benefit from low-rank or block-sparse approximations, fused projections, and optimized GEMM kernels to improve arithmetic intensity. Algorithmic modifications including a reduction in the number of ET layers, cutoff smoothing, or lightweight geometric encoders may help maintain accuracy while improving scaling for larger molecular systems.
Looking ahead, GPUs targeting MLFF workloads should support larger configurable L1/shared caches, hardware prefetching for non-contiguous memory access, tensor cores optimized for small size GEMMs, low-latency kernel launch and fusion and adaptive warp scheduling for heterogeneous neighborhoods. High bandwidth interconnects between compute units would minimize data transfer costs.

\section{Conclusions}
MLFFs are emerging workloads used in MD simulations due to the accuracy improvement associated with these force fields. However, this comes with an added computational overhead compared to the CFFs. This paper introduces poly-alanine slices as a molecular system of controllable system size to analyze the effect of MLFFs in GPUs with scaling workloads. The analysis covers two popular types of MLFF network architectures (ANi2X and TorchMD-Net), and their performance with an increasing size workload.

The introduction of the workload and detailed profiling strategies exposed the performance bottlenecks in MLFFs. The presence of forward and backward path calculations at every time step significantly differentiates MLFF calculation from the conventional neural network inference, hence introducing unique memory and computational challenges. Furthermore, the dynamic neighborhood changes in the atomic environment also create memory access challenges in AEV/ET calculations compared to conventional embedding techniques used in ML, which necessitates new memory architectural designs for MD related workloads that are not necessary for standard ML workloads. Furthermore, fused kernels and introducing parallelization techniques like domain decomposition can contribute in effectively utilizing the existing GPU infrastructure.

\section*{Acknowledgment}
This research is funded by the imec.prospect project IMPACT. The authors thank Peter Kourzanov for valuable feedback and troubleshooting assistance, Geert Vanmeerbeeck ,Tom Vander Aa for technical support, and Peter Vrancx and Eleni Litsa for insightful discussions.
\bibliographystyle{IEEEtranS}
\bibliography{reference}
\vspace{12pt}
\end{document}